\documentclass[aps,twocolumn,prb,showpacs]{revtex4-1}
\usepackage{graphicx}
\usepackage{amssymb,amsmath}
\usepackage{xcolor}

\newcommand{\bk}{\mathbf{k}}

\begin{document}

\title{Topological Properties of Electrons in Honeycomb Lattice with Kekul\'{e} Hopping Textures}

\author{Long-Hua Wu}
\email{Wu.Longhua@nims.go.jp}
\affiliation{$^1$International Center for Materials Nanoarchitectonics
(WPI-MANA), National Institute for Materials Science, Tsukuba 305-0044, Japan \\ 
$^2$Graduate School of Pure and Applied Sciences, University of Tsukuba,
Tsukuba 305-8571, Japan}
\author{Xiao Hu}
\email{Hu.Xiao@nims.go.jp}
\affiliation{$^1$International Center for Materials Nanoarchitectonics
(WPI-MANA), National Institute for Materials Science, Tsukuba 305-0044, Japan \\ 
$^2$Graduate School of Pure and Applied Sciences, University of Tsukuba,
Tsukuba 305-8571, Japan}

\date{\today}
\pacs{03.65.Vf,73.43.-f,73.23.-b}

\begin{abstract}
Honeycomb lattice can support electronic states exhibiting Dirac energy
dispersion, with graphene as the icon. We propose to derive nontrivial
topology by grouping six neighboring sites of honeycomb lattice into hexagons
and enhancing the inter-hexagon hopping energies over the intra-hexagon ones.
We reveal that this manipulation opens a gap in the energy dispersion and
drives the system into a topological state.  The nontrivial topology is
characterized by the $\mathbb{Z}_2$ index associated with a pseudo
time-reversal symmetry emerging from the $C_6$ symmetry of the Kekul\'{e}
hopping texture, where the angular momentum of orbitals accommodated on the
hexagonal "artificial atoms" behaves as the pseudospin. The size of
topological gap is proportional to the hopping-integral difference, which can
be larger than typical spin-orbit couplings by orders of magnitude and
potentially renders topological electronic transports available at high
temperatures.
\end{abstract}

\maketitle

\section{Introduction}
Honeycomb lattice can host electrons with Dirac-like linear
dispersion due to its $C_3$ crystal symmetry~\cite{Wallace1947}, and interests
in questing for systems with honeycomb lattice structure flourished since the
discovery of graphene produced by the scotch-tape
technique~\cite{Novoselov2004,GrapheneBook,Geim2009}. The Dirac dispersion and
the associated chiral property of electronic wave functions accommodated on honeycomb
lattice make it an ideal platform for exploring topological states~\cite{QHE,TKNN} without
external magnetic field. It was
shown first that a quantum anomalous Hall effect (QAHE) can be realized when
complex hopping integrals among next-nearest-neighboring sites of honeycomb
lattice are taken into account~\cite{Haldane1988}. Later on it was revealed
that the intrinsic spin-orbit coupling (SOC) in honeycomb lattice can provide this complex
hopping integrals, which drives spinful electrons into a topological state with preserved
time-reversal (TR) symmetry, known as quantum spin Hall effect
(QSHE)~\cite{KaneRMP,ZhangRMP,Kane2005a,Kane2005b}. Quite a number of activities have been
devoted towards realizing topological states in electron systems on honeycomb
lattice, such as QAHE by straining~\cite{GeimNPhys2010,Gomes2012}, twisting
\cite{Hunt2013,He2013} and decorating graphene~\cite{QNiuPRB2010}, and QAHE
with spin-polarized edge currents in terms of the antiferromagnetic exchange field
and staggered electric potential~\cite{LiangNJP2013}. Honeycomb lattice can
also be tuned to support topological states in photonic
crystals~\cite{Haldane2008,Wu2015} and cold atoms~\cite{Tarruell2012}.

\begin{figure}[t]
  \centering
  \includegraphics[width=.8\linewidth]{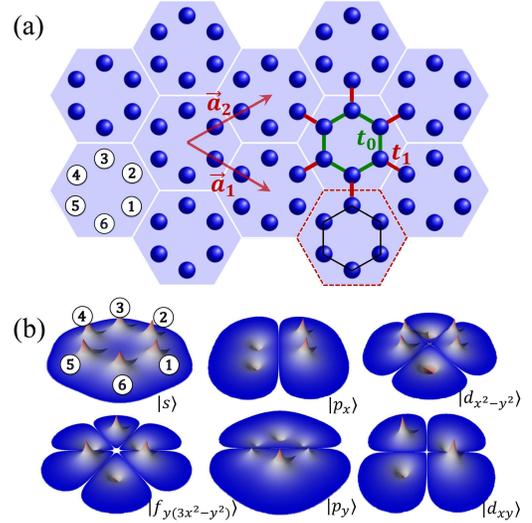}
  \caption{(Color online) (a) Honeycomb lattice with a Kekul\'{e} texture in
  hopping energies between NN sites: $t_0$ inside hexagons as denoted by the
  green bonds and $t_1$ between hexagons by red ones. The Red dashed hexagon
  is the primitive cell of triangular lattice with lattice vectors
  $\vec{a}_1$, $\vec{a}_2$ and lattice constant $a_0=|\vec{a}_1|=|\vec{a}_1|$.
Numbers $1,\dots,6$ in circle index atomic sites within a hexagon. (b)
Emergent orbitals in the hexagonal artificial atom.}
  \label{model}
\end{figure}

In the present work, we explore possible topological properties in honeycomb
lattice by introducing a Kekul\'{e} texture in hopping energy between \textit{
nearest-neighboring} (NN) sites. We take a hexagonal primitive unit cell and
view the honeycomb lattice as a triangle lattice of hexagons [see the dashed
red line in Fig.~\ref{model}(a)]. When the inter-hexagon hopping $t_1$ is
tuned to be larger than the intra-hexagon one $t_0$, a topological gap is
opened at the $\Gamma$ point accompanied by a band inversion between orbitals
with opposite spatial parities accommodated on hexagons [see
Fig.~\ref{model}(b)]. A pseudo-TR symmetry associated with a pseudospin
degree of freedom and Kramers doubling in the emergent orbitals are revealed
based on $C_6$ point group symmetry, which generates the $\mathbb{Z}_2$
topology. For experimental implementations, we discuss that, along with many
other possibilities, the \textit{molecular graphene} realized by placing
carbon monoxides (CO) periodically on Cu [111] surface~\cite{Gomes2012} is a
very promising platform to realize the present idea, where the Kekul\'{e}
texture can be controlled by adding extra CO molecules.

The paper is organized as follows. In Sec.~\ref{SecModel}, we start with a
$C_6$ symmetric Kekul\'{e} hopping texture among NN sites in honeycomb
lattice, where a pseudo-TR symmetry and pseudospin emerge. In
Sec.~\ref{SecTopo}, we derive an effective low-energy Hamiltonian of the
system and reveal the $\mathbb{Z}_2$ topology. The topological properties
of the system is demonstrated by the edge state and associated Hall
and longitudinal conductances in sections ~\ref{SecEdge} and \ref{SecSOC}. We
discuss a promising experimental platform for realizing the topological state
in Sec.~\ref{SecExp}. Concluding remarks are presented in
Sec.~\ref{SecConclude}.

\section{Kekul\'{e} hopping texture and emergent orbitals}
\label{SecModel}
We start from a spinless tight-binding Hamiltonian on honeycomb lattice
\begin{equation}
  H= \varepsilon_0 \sum_i c_i^\dagger c_i + t_0\sum_{\langle i,j\rangle}
  c_i^\dagger c_j+ t_1\sum_{\langle i', j'\rangle}  c_{i'}^\dagger c_{j'},
  \label{tbmodel}
\end{equation}
with $c_i$ the annihilation operator of electron at atomic site $i$ with
on-site energy $\varepsilon_0$ satisfying anti-commutation
relations, $\langle i,j\rangle$ and $\langle i',j'\rangle$ running over NN
sites inside and between hexagonal unit cells with hopping energies $t_0$
and $t_1$ respectively [see Fig.~\ref{model}(a)]. The orbitals are
considered to be the simplest one without any internal structure, same as the
$\pi$ electron of graphene. Below we are going to detune the hopping energy
$t_1$ while keeping $t_0$ constant, and elucidate possible changes in the
electronic state.  In this case, the pristine honeycomb lattice of
individual atomic sites is better to be considered as a triangular lattice of
hexagons, with the latter characterized by $C_6$ symmetry.

Let us start with the Hamiltonian within a single hexagonal unit cell
\begin{equation}
  H_0 = t_0\Psi^\dagger
  \left(
  \begin{array}{cccccc}
    0 & 1 & 0 & 0 & 0 & 1 \\
    1 & 0 & 1 & 0 & 0 & 0 \\
    0 & 1 & 0 & 1 & 0 & 0 \\
    0 & 0 & 1 & 0 & 1 & 0 \\
    0 & 0 & 0 & 1 & 0 & 1 \\
    1 & 0 & 0 & 0 & 1 & 0
  \end{array}
  \right) \Psi,
  \label{eq:t0}
\end{equation}
where $\Psi = [c^1,c^2,c^3,c^4,c^5,c^6]^T$ [see Fig.~\ref{model}(a)].
The eigenstates of Hamiltonian $H_0$ are given by
\begin{eqnarray}
&&|s\rangle = [1,1,1,1,1,1]^T;\nonumber \\
&&|p_x\rangle = [1,1,0,-1,-1,0]^T;\nonumber \\
&&|p_y\rangle  = [1,-1,-2,-1,1,2]^T; \nonumber \\
&&|d_{x^2-y^2}\rangle = [1,1,-2,1,1,-2]^T; \nonumber \\
&&|d_{xy}\rangle = [1,-1,0,1,-1,0]^T; \nonumber \\
&&|f_{y(3x^2-y^2)}\rangle = [1,-1,1,-1,1,-1]^T
\label{spdf}
\end{eqnarray}
with eigenenergies $2t_0,t_0,t_0,-t_0,-t_0$ and $-2t_0$ respectively,
up to normalization factors. As shown in Fig.~\ref{model}(b), the
\textit{emergent} orbitals accommodated on the hexagonal ``artificial atom'' take the
shapes similar to the conventional $s$, $p$, $d$ and $f$ atomic orbitals in solids.

A pseudo-TR symmetry operator can be composed in the present system with $C_6$
symmetry: $\mathcal{T}=\mathcal{UK}$ with $\mathcal{K}$ the complex conjugate
operator and $\mathcal{U}=-i\sigma_y$, where $\sigma_y$ is the Pauli matrix.
It can be checked straightforwardly that $\mathcal{U}$ corresponds to $\pi/2$
rotation for $p$ orbitals and $\pi/4$ rotation for $d$ orbitals given in
Eq.~(\ref{spdf}), which yields $\mathcal{U}^2=-1$ in the space formed by the
$p$ and $d$ orbitals \cite{Wu2015}.  Therefore, the pseudo-TR symmetry
satisfies the relation $\mathcal{T}^2=-1$, which is same as that for fermionic
particles even though the spin degree of freedom of electron has not been
considered here. This indicates that electrons acquire a new
\textit{pseudospin} degree of freedom in the present system as far 
as the low-energy physics is concerned.

Explicitly the wave functions carrying pseudospins are given by the
emergent orbitals with eigen angular momentum
\begin{equation}
  |p_\pm\rangle = \frac{1}{\sqrt{2}}\left(|p_x\rangle \pm i |p_y\rangle\right);
  |d_\pm\rangle = \frac{1}{\sqrt{2}}\left(|d_{x^2-y^2}\rangle \pm i |d_{xy}\rangle\right).
  \label{pspin}
\end{equation}
Distinguished from the intrinsic spin, the pseudospin is
directly related to the chiral current density on the hexagon. For a lattice
model, the current density between two sites is given by
$I_{jk}=(i/\hbar)[t_0c_j^\dagger c_k -t_0^* c_k^\dagger c_j]$.
The current distributions evaluated using wave functions in Eq.~(\ref{pspin})
for the pseudospin-up and -down states are shown in Figs.~\ref{bulkband}(a)
and (b) with anticlockwisely and clockwisely circulating currents.  By
considering the hexagonal artificial atoms composed by six sites in honeycomb
lattice, one harvests states with angular momentum merely from simple
orbitals, such as $\pi$ electrons in graphene. The pseudo-TR symmetry is
supported by the $C_6$ crystal symmetry, sharing the same underlying physics
with the topological crystalline insulator~\cite{FuPRL2011}. However,
crystal-symmetry-protected topological insulators addressed so far need strong
SOCs to achieve band inversions~\cite{Hsieh2012,Dziawa2012,Xu2012}, which is
different from the present approach as revealed below.

\section{Topological phase transition}
\label{SecTopo}
\begin{figure}[t]
  \centering
  \includegraphics[width=\linewidth]{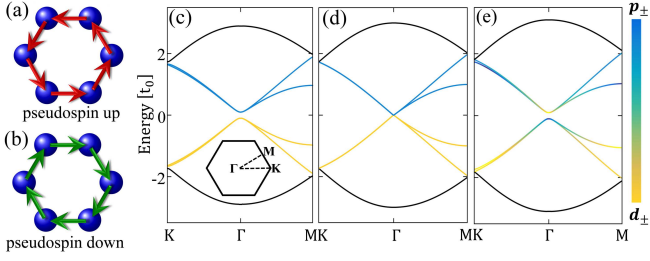}
  \caption{(Color online) (a) and (b) Current densities in the pseudospin-up channel
  ($p_+$ or $d_+$) and pseudospin-down channel ($p_-$ or $d_-$) respectively.
  Band dispersions for the system given in Fig.~\ref{model}: (c) $t_1 =
  0.9t_0$ (Inset: Brillouin zone of the triangular lattice), (d) $t_1 = t_0$
  and (e) $t_1 = 1.1t_0$. Blue and yellow are for $|p_\pm\rangle$ and
  $|d_\pm\rangle$ orbitals respectively, and rainbow for hybridization between
  them. The on-site energy is taken $\varepsilon_0=0$.}
  \label{bulkband}
\end{figure}

We calculate the energy dispersion of Eq.~(\ref{tbmodel}) for several
typical values of $t_1$ (hereafter the on-site energy is put as
$\varepsilon_0=0$ without losing generality). As shown in Fig.~\ref{bulkband},
there are two two-fold degeneracies at the $\Gamma$ point corresponding to the
two two-dimensional (2D) irreducible representations of $C_6$ point group.
Projecting the wave functions for $t_1 = 0.9t_0$ onto the orbitals given in
Fig.~\ref{model}(b), it is found that the topmost two valance bands show the
character of $d$ orbitals whereas the lowest two conduction bands behave like
$p$ orbitals [see Fig.~\ref{bulkband}(c)], with the order in energy same as
those listed in Eq.~(\ref{spdf}). For $t_1 = t_0$, the $d$ and $p$ bands become degenerate at the
$\Gamma$ point and double Dirac cones appear [see Fig.~\ref{bulkband}(d)],
which are equivalent to the ones at
$K$ and $K'$ points in the unfolded Brillouin zone of honeycomb lattice with
the rhombic unit cell of two sites. When $t_1$ increases further from
$t_0$, a band gap reopens at the $\Gamma$ point. As shown in Fig.~\ref{bulkband}(e)
for $t_1=1.1t_0$, the valence (conduction) bands are now occupied by $p$ ($d$) orbitals
around the $\Gamma$ point,
opposite to the order away from the $\Gamma$ point, and to that before gap
closing. Therefore, a band inversion between $p$ and $d$ orbitals takes place at
the $\Gamma$ point when the inter-hexagon hopping energy is increased across
the topological transition point $t_1=t_0$, namely the pristine honeycomb lattice.

We can characterize the topological property of the gap-opening
transition shown in Fig.~\ref{bulkband} by a low-energy effective Hamiltonian
around the $\Gamma$ point. Since
the bands near Fermi level are predominated by $p$ and $d$ orbitals, it is
sufficient to downfold the six-dimensional Hamiltonian $H(\mathbf{k})$
associated with the tight-binding model (\ref{tbmodel}) into the
four-dimensional subspace $[p_+,d_+,p_-,d_-]$. 
The second term in Eq.~(\ref{tbmodel}) is then simply given by
\begin{equation}
  h_0' =
  \left(
  \begin{array}{cccc}
   t_0&  0  & 0 & 0 \\
    0 &-t_0 & 0 & 0 \\
    0 &  0  &t_0& 0 \\
    0 &  0  & 0 &-t_0
  \end{array}
  \right),
  \label{hinside}
\end{equation}
in the four-dimensional subspace. Contributions from the third term in
Eq.~(\ref{tbmodel}) to the effective Hamiltonian should be evaluated
perturbatively. First, we list the inter-hexagon hoppings in terms of
6$\times$6 matrices $h_1, h_2, h_3, h_1^\dagger, h_2^\dagger$ and
$h_3^\dagger$ with
\[
  h_1 =
  \left(
  \begin{array}{cccccc}
    0 & 0 & 0 &t_1 & 0 & 0 \\
    0 & 0 & 0 & 0 & 0 & 0 \\
    0 & 0 & 0 & 0 & 0 & 0 \\
    0 & 0 & 0 & 0 & 0 & 0 \\
    0 & 0 & 0 & 0 & 0 & 0 \\
    0 & 0 & 0 & 0 & 0 & 0 \\
  \end{array}
  \right),
  h_2 =
  \left(
  \begin{array}{cccccc}
    0 & 0 & 0 & 0 & 0 & 0 \\
    0 & 0 & 0 & 0 & t_1 & 0 \\
    0 & 0 & 0 & 0 & 0 & 0 \\
    0 & 0 & 0 & 0 & 0 & 0 \\
    0 & 0 & 0 & 0 & 0 & 0 \\
    0 & 0 & 0 & 0 & 0 & 0 \\
  \end{array}
  \right),
\]
\begin{equation}
  h_3 =
  \left(
  \begin{array}{cccccc}
    0 & 0 & 0 & 0 & 0 & 0 \\
    0 & 0 & 0 & 0 & 0 & 0 \\
    0 & 0 & 0 & 0 & 0 & t_1 \\
    0 & 0 & 0 & 0 & 0 & 0 \\
    0 & 0 & 0 & 0 & 0 & 0 \\
    0 & 0 & 0 & 0 & 0 & 0 \\
  \end{array}
  \right)
  \nonumber
\end{equation}
on the basis of $[c^1,c^2,c^3,c^4,c^5,c^6]$.  Following the standard
procedures~\cite{Sakurai}, they can be projected to the subspace spanned
by $[p_+,d_+,p_-,d_-]$
\begin{eqnarray}
  h_1'&=&
  \frac{t_1}{12}
  \left(
  \begin{array}{cccc}
    -2& \sqrt{3}-i  & -1-\sqrt{3}i & 2i \\
    -\sqrt{3}-i & 2 & -2i & -1+\sqrt{3}i \\
    -1+\sqrt{3}i & -2i &-2& \sqrt{3}+i \\
    2i & -1-\sqrt{3}i & -\sqrt{3}+i & 2
  \end{array}
  \right), \nonumber \\
  h_2'&=& h_1'^*, \nonumber \\
  h_3'&=&
  \frac{t_1}{6}
    \left(
    \begin{array}{cccc}
      -1 & i & 1 & i \\
       i & 1 &-i & 1 \\
       1 &-i &-1 & -i \\
       i & 1 &-i & 1
    \end{array}
    \right).
    \label{hinter}
\end{eqnarray}

With Fourier transformations of matrices in Eqs.~(\ref{hinside}) and
(\ref{hinter}), one obtains the effective low-energy Hamiltonian
$H'(\mathbf{k})$ on the basis $[p_+,d_+,p_-,d_-]$ in the momentum space.
Expanding $H'(\mathbf{k})$ around the $\Gamma$ point up to the lowest-orders
of $\bk$, one arrives at
\begin{equation}
  H'(\bk\rightarrow\Gamma) =
  \left(
  \begin{array}{cc}
    H_+(\bk) & \mathbf{0} \\
    \mathbf{0} & H_-(\bk) \\
  \end{array}
  \right)
  \label{hgamma}
\end{equation}
with
\begin{eqnarray}
  H_+(\bk) &=&
  \left(
  \begin{array}{cc}
    -\delta t + \frac{1}{2} a_0^2t_1\bk^2 & \frac{i}{2}a_0t_1k_+ \\
    -\frac{i}{2}a_0t_1k_- & \delta t - \frac{1}{2}a_0^2t_1 \bk^2
  \end{array}
  \right), \nonumber \\
  H_-(\bk) &=&
  \left(
  \begin{array}{cc}
    -\delta t + \frac{1}{2}a_0^2t_1 \bk^2 & \frac{i}{2}a_0t_1k_-\\
    -\frac{i}{2}a_0t_1k_+ & \delta t - \frac{1}{2}a_0^2t_1\bk^2
  \end{array}
  \right), 
  \label{Hkp}
\end{eqnarray}
where $\delta t = t_1 - t_0$, $\bk=(k_x,k_y)$, $k_\pm = k_x\pm i k_y$, $\mathbf{0}$ is a
$2\times2$ zero matrix, and $a_0$ is the lattice constant of the triangular
lattice. For $\delta t=0$, the Hamiltonians $H_+(\mathbf{k})$ and
$H_-(\mathbf{k})$ in Eq.~(\ref{Hkp}) are the same as the well-known one for
honeycomb lattice around the $K$ and $K'$ points \cite{NetoRMP2009}, where the
quadratic terms of momentum in the diagonal parts can be neglected.

For $\delta t>0$, however, the quadratic terms are crucially important
since they induce a band inversion~\cite{Bernevig2006}, resulting in
the orbital hybridization in the band structures denoted by
the rainbow colors in Fig.~\ref{bulkband}(e). Associated with a skyrmion in
the momentum space for the orbital distributions in the individual pseudospin
channels, a topological state appears characterized by the $\mathbb{Z}_2$
topological invariant~\cite{Kane2005a,Kane2005b,Wu2015,FuPRB2007}. It is clear
that for $\delta t<0$ there is no band inversion taking place and thus the
band gap is trivial as shown in Fig.~\ref{bulkband}(c).

It is worthy noticing that, comparing with the Kane-Mele model for the
honeycomb lattice \cite{Kane2005a,Kane2005b}, the mass term $\delta t (>0)$ in
Eq.~(\ref{Hkp}) can be considered as an effective SOC associated with the
pseudospin, namely $\lambda_{\rm eSOC} = \delta t$.  For $\delta t=0.1t_0$, a
moderate Kakul\'{e} texture in hopping energy, the effective SOC is
approximately 3000 times larger than the \textit{real} SOC in magnitude in
graphene where $\lambda_{\rm SOC}\simeq$~0.1meV and $t_0=2.7$eV.  The huge
effective SOC is due to its pure electronic character as compared with the
intrinsic SOC originated from the relativistic effect. This is one of the
fantastic aspects of the present approach, which renders a topological gap
corresponding to temperature of thousands of Kelvin.

\begin{figure}[t]
  \centering
  \includegraphics[width=\linewidth]{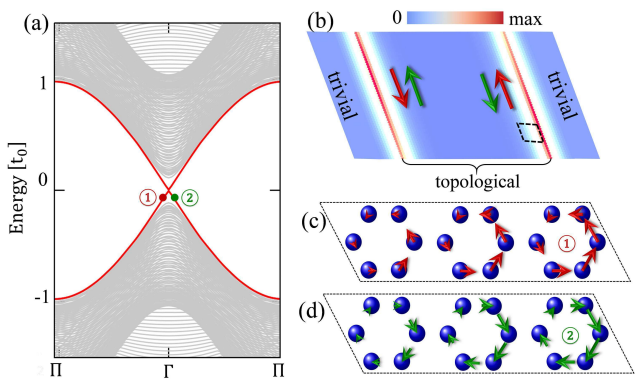}
  \caption{(Color online) (a) Band dispersion of a ribbon system of 36 hexagons with
  $t_1=1.1t_0$ cladded from both sides by 10 hexagons with $t_1 = 0.9t_0$. (b)
  Real-space distribution of the in-gap states associated with the red solid
  dispersion curves in (a).  (c) and (d) Real-space distributions of current
  densities in pseudospin up and down channels at the momenta indicated by the
  red and green dots 1 and 2 in (a) within the rhombic area sketched by dashed
  line in (b); the excess currents in pseudospin up and down channels are
  indicated by red and green arrows in (b).}
  \label{edge}
\end{figure}

\section{Topological edge states and associated conductances}
\label{SecEdge}
\subsection{Topological edge states}
In order to check the edge state in the present system, we
consider a ribbon of hexagonal unit cells of $t_1 = 1.1t_0$ with its two edges
cladded by hexagonal unit cells of $t_1 = 0.9t_0$. As can be seen in
Fig.~\ref{edge}(a), additional states appear in the bulk gap as indicated by
the red solid curves carrying double degeneracy. Plotting the spatial
distribution of the corresponding wave functions, we find
that the in-gap states are localized at the two interfaces between topological
and trivial regions [see Fig.~\ref{edge}(b)]. As displayed in
Fig.~\ref{edge}(c) [(d)], there is an excess upward (downward) edge current in the
pseudospin-up (-down) channel associated with the state indicated by the
red (green) dot in Fig.~\ref{edge}(a).

\begin{figure}[t]
  \centering
  \includegraphics[width=.8\linewidth]{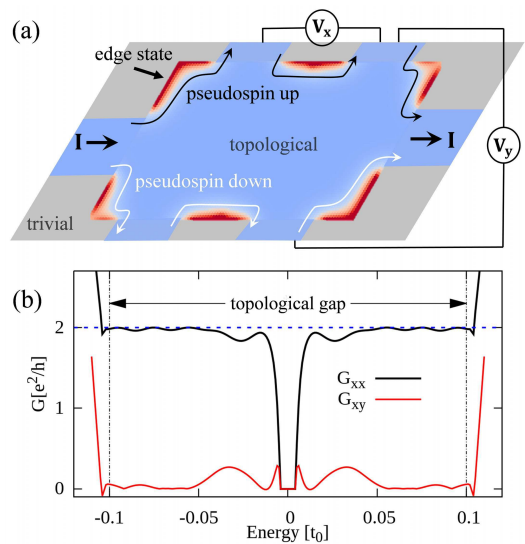}
  \caption{(Color online) (a) Schematic configuration of a six-terminal Hall bar where a topological sample
  (light blue region) with $t_1 = 1.1t_0$ is embedded in a trivial environment
  (gray region) with $t_1 = 0.9t_0$.  The size of topological scattering
  region is $240a_0\times120a_0$, and the width of each semi-infinite lead is
  $40a_0$. The injected current flows along the edges of topological sample as
  indicated by the red parts between electrodes. (b) Longitudinal and Hall
  conductances of the Hall bar as a function of energy of incident electrons.
  The on-site energy is taken $\varepsilon_0=0$. A rhombic topological sample
  is taken for ease of calculation.}
  \label{hall}
\end{figure}

\subsection{Hall and longitudinal conductances}
At the interface between topological and trivial regimes, the crystal symmetry
is reduced from $C_6$ to $C_2$, which breaks the pseudo-TR symmetry in
contrast to the real TR symmetry. As the result, a mini gap of $\sim0.01t_0$
[unnoticeable in the scale of Fig.~\ref{edge}(a)] opens in the otherwise
gapless edge state at the $\Gamma$ point due to the coupling between two
pseudospin channels. In order to quantitatively check possible
backscatterings caused by this mini gap, we perform calculations on the
longitudinal and Hall conductances based on a Hall bar system as sketched in
Fig.~\ref{hall}(a). It is clear that the current $I$ injected from the left
electrode divides itself into two branches according to the pseudospin states,
namely pseudospin up (down) electrons can flow only in the upper (lower) edge
of the Hall bar. By matching wave functions at the interfaces between the six
semi-infinite electrodes and the topological scattering region
\cite{AndoPRB1991,GrothNJP2014}, one can evaluate the transmissions of plane
waves scattered among all the six leads, and then the longitudinal and Hall
conductances, $G_{xx} = \rho_{xx}/(\rho_{xx}^2+\rho_{xy}^2)$ and $G_{xy} =
\rho_{xy}/(\rho_{xx}^2+\rho_{xy}^2)$ respectively, by the
Landauer-B\"{u}ttiker formalism~\cite{Landauer1999}, where $\rho_{xx}$ and
$\rho_{xy}$ are the longitudinal and transverse resistances respectively.
Similar to the case of QSHE with magnetic impurities~\cite{TkachovPRL2010}, in
the present system the values of conductivity deviate from the quantized ones
when Fermi level falls in the mini gap of $\sim 0.01t_0$ as shown in
Fig.~\ref{hall}(b). It is noticed, however, that both $G_{xx}$ and $G_{xy}$
heal quickly after several periods of oscillations that come from
interferences between the two pseudospin channels.  It is emphasized that
almost perfectly quantized conductances $G_{xx}= 2e^2/h$ and
$G_{xy}=0$~\cite{Bernevig2006,Konig2007} are achieved for Fermi level beyond
$0.04t_0$ up to the bulk gap edge at $0.1t_0$, where the edge states with
almost perfect linear dispersions hardly feel the existence of the mini gap
and essentially no appreciable backscattering exists. 

\begin{figure}[t]
  \centering
  \includegraphics[width=.85\linewidth]{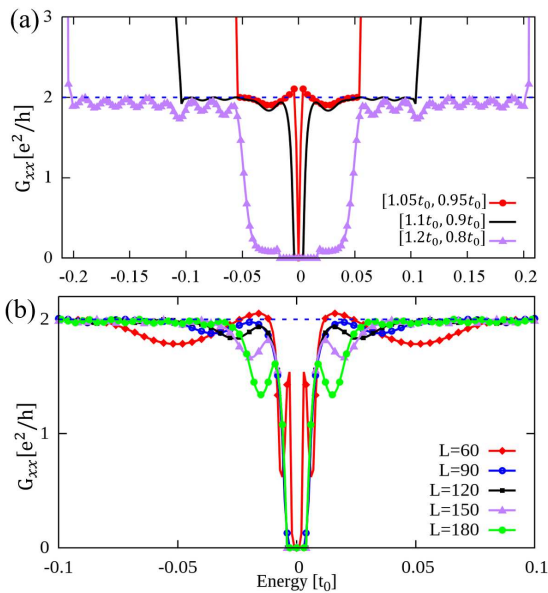}
  \caption{(Color online) Longitudinal conductance $G_{xx}$ of the topological sample given
  in Fig.~\ref{hall}(a) as a function of the energy of
  injected electrons: (a) for several typical values of inter-hexagon hopping
  integrals $[1.05t_0,0.95t_0]$, $[1.1t_0,0.9t_0]$ and $[1.2t_0,0.8t_0]$,
  where the first (second) inside bracket is for the topological (trivial)
  region; (b) with several typical system sizes $2L \vec{a}_1\times L
  \vec{a}_2$, where the width of the electrodes is fixed at $40 a_0$ and the
  inter-hexagon hopping integral is fixed at $t_1=1.1t_0$ and $t_1=0.9t_0$ for
  the topological and trivial regions respectively.}
  \label{Shall}
\end{figure}

Now we discuss the hopping integral dependence of the longitudinal conductance.
The size of scattering region is same as in Fig.~\ref{hall}(a) and fixed for
all cases. As displayed in Fig.~\ref{Shall}(a), $G_{xx}$ saturates to the
quantized value $2e^2/h$ as expected for a $\mathbb{Z}_2$ topological state
for all the cases with $t_1 = 1.05t_0, 1.1t_0$ and $1.2t_0$ in the topological
region (whereas $0.95t_0, 0.9t_0$ and $0.8t_0$ in the trivial region
correspondingly) when Fermi level is set away from the mini gaps, accompanied
by oscillations due to interferences between two pseudospin channels.

Here we show the sample size dependence of the longitudinal conductance.  We
fix inter-hexagon hopping integrals at $1.1t_0$ and $0.9t_0$ in the
topological and trivial regions respectively. As displayed in
Fig.~\ref{Shall}(b), $G_{xx}$ saturates in all cases to the quantized value
$2e^2/h$ when Fermi level is shifted away from the mini gap. The topological
edge transport remain unchanged when the topological sample becomes large.

\section{Real spin and QSHE}
\label{SecSOC}
In addition to the pseudospin, the \textit{real} spin degree of freedom also
contributes to transport properties in realistic systems. In absence of the
real SOC, the results presented in Fig.~\ref{hall} remain exactly the same,
with an additional double degeneracy due to the two real spin and thus
$G_{xx}=4e^2/h$.

\begin{figure}[t]
  \centering
  \includegraphics[width=.95\linewidth]{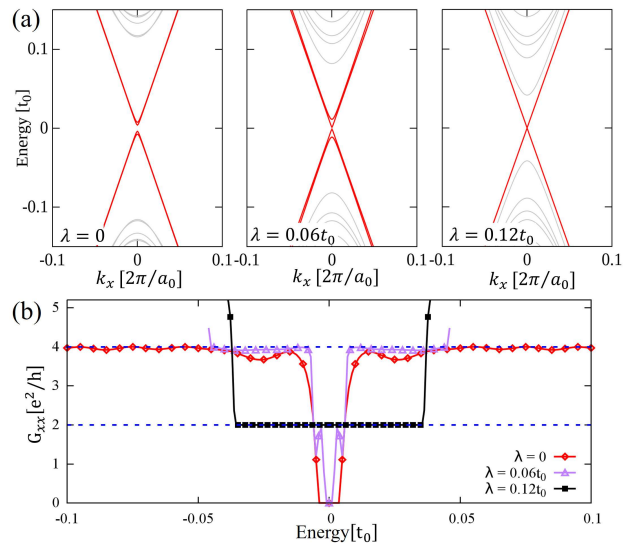}
  \caption{(Color online) (a) Dispersion relations and (b) longitudinal conductances of the
  topological system same as that given in Fig.~\ref{hall}(a) except that
  finite SOC is included.}
  \label{HallSOC}
\end{figure}

An intrinsic SOC is induced when next-nearest-neighbor hoppings in
honeycomb lattice are taken into account~\cite{Kane2005a,Kane2005b}. The low-energy
Hamiltonian around the $\Gamma$ point in Eq.~(\ref{hgamma}) is
then modified to
\begin{equation}
   \tilde{H}(\bk) =
  \left(
  \begin{array}{cc}
     \tilde{H}_+(\bk) & \mathbf{0} \\
    \mathbf{0} &  \tilde{H}_-(\bk) \\
  \end{array}
  \right)
\end{equation}
with
\begin{eqnarray}
  \tilde{H}_+(\bk) &=&
  \left(
  \begin{array}{cc}
    -\delta t - \nu\lambda + \frac{1}{2} a_0^2t_1 \bk^2 & \frac{i}{2}a_0t_1k_+ \\
    -\frac{i}{2}a_0t_1k_- & \delta t + \nu\lambda - \frac{1}{2}a_0^2t_1 \bk^2
  \end{array}
  \right),  \nonumber \\
  \tilde{H}_-(\bk) &=&
  \left(
  \begin{array}{cc}
    -\delta t + \nu\lambda + \frac{1}{2} a_0^2t_1 \bk^2 & \frac{i}{2}a_0t_1k_- \\
    -\frac{i}{2} a_0t_1k_+ & \delta t -\nu\lambda - \frac{1}{2}a_0^2t_1 \bk^2
  \end{array}
  \right), \nonumber
\end{eqnarray}
where $\nu = 1$ and $-1$ stand for real spin-up and -down states respectively.
Therefore, for up real spin SOC enhances (suppresses) the topological gap in
the pseudospin up (down) channel presuming $\lambda>0$ [see the left and
central panels of Fig.~\ref{HallSOC}(a)]. As far as $\lambda<\delta t$, the
system remains the $\mathbb{Z}_2$ topological state associated with the
pseudospin, where electrons with up pseudospin and down pseudospin counter
propagate at edges, both carrying on up and down real spins. The longitudinal
conductance $G_{xx}$ saturates to $4e^2/h$ as displayed in
Fig.~\ref{HallSOC}(b).

When SOC is increased to $\lambda=\delta t$, the pseudospin down (up) channel
with real spin up (down) is driven into a semi-metallic state with zero band
gap and the Dirac dispersion appears at the $\Gamma$ point.  When SOC goes
beyond $\delta t$, this Dirac dispersion opens a gap accompanied by a
topological phase transition. The system now takes a QSHE state where at edges
electrons with up real spin and pseudospin propagate oppositely to electrons
with down real spin and pseudospin. Evaluating the longitudinal conductance,
one finds that $G_{xx}$ is quantized exactly to $2e^2/h$ (see
Fig.~\ref{HallSOC}(b)), and as shown in the right panel of Fig.~\ref{HallSOC}(a)
there is no mini gap in the edge states, as protected by \textit{real}
time-reversal symmetry~\cite{Kane2005a,Kane2005b}.

\begin{figure}[t]
  \centering
  \includegraphics[width=\linewidth]{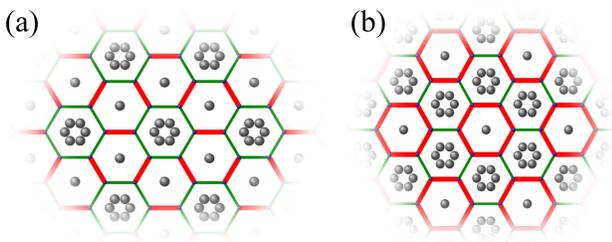}
  \caption{(Color online) Molecular graphene realized by decorating the Cu
  [111] surface with a triangular lattice of CO molecules: (a) $t_1 > t_0$
  generating topological state, (b) $t_1< t_0$ for trivial state. Gray balls
  are CO molecules decorated by STM techniques, and red thick bonds are
  shorter than green thin ones which generates the Kekul\'{e} hopping
  textures.}
  \label{devices}
\end{figure}

\section{Discussions}
\label{SecExp}
Much effort has been devoted towards realizing the Dirac-like energy dispersion in
artificial honeycomb lattices \cite{Polini2013}, ranging from optical lattices
\cite{Tarruell2012,Wunsch2008} to 2D electron gases modulated by periodic
potentials \cite{Gomes2012,GibertiniPRB2009,SGLouie2009}. All these systems
provide promising platforms for realizing topological properties by detuning
effective hopping energy among NN sites either by
modulating muffin-tin potentials or bond lengthes periodically. To be specific,
here we focus on how to achieve a topological state on the Cu [111] surface
modulated by triangular gates of carbon monoxide (CO)
molecules~\cite{Gomes2012}. When extra CO molecules are placed at specific
positions over the pristine molecular graphene, the bonds of the hexagons
surrounding them are elongated since the CO clusters enhance local
repulsive potentials and push electrons away from them, which reduces the
corresponding electron hopping energies~\cite{SGLouie2009}. It is extremely
interesting from the present point of view that Kekul\'{e} hopping textures
have already been achieved in experiments~\cite{Gomes2012}.  We propose to
place extra CO atoms in the pattern displayed in Fig.~\ref{devices}(a), where
the intra-hexagon hopping energy $t_0$ (green thin bonds) surrounding the CO
clusters is reduced and the inter-hexagon hopping energy $t_1$ (red thick
bonds) is enhanced relatively. According to the above discussions, the system
displayed in Fig.~\ref{devices}(a) with $t_1>t_0$ should take a topological state. The
Kekul\'{e} hopping texture in Fig.~\ref{devices}(b), dual to that shown in
Fig.~\ref{devices}(a), was realized in experiments~\cite{Gomes2012}, where
the system takes a topologically trivial state because $t_1<t_0$.

The underlying idea of the present scheme for achieving the $\mathbb{Z}_2$ topological state
is to create artificial orbitals carrying on opposite orbital angular
momenta and parities with respect to spatial inversion symmetry, and to
induce a band inversion between them by introducing a Kekul\'{e} hopping
texture on honeycomb lattice. In the sense that it does not require SOC, the present state
may be understood as a quantum \textit{orbital} Hall effect. The
topological properties can also be extended to photonic crystals~\cite{Wu2015},
cold atoms, and even phonon systems where sound waves form band due to
periodic configurations elastic materials.

\section{Conclusion}
\label{SecConclude}
We propose to derive topological properties by
modulating electron hopping energy between nearest-neighbor sites of honeycomb
lattice. Due to the Kekul\'{e} hopping texture with $C_6$ symmetry,
atomic-like orbitals emerge, which carry a pseudospin degree of freedom characterizing
a pseudo time-reversal symmetry and Kramers doubling. We reveal that the effective
spin-orbit coupling associated with the pseudospin degree of freedom can be larger than the
intrinsic one by orders of magnitude because of the pure electronic
origin. The present work offers a new possibility for achieving topological
properties and related novel quantum properties and functionalities at high
temperatures.

\begin{acknowledgments}
The authors thank Y.-Y. Wang, L.~You, Z.-F.~Xu, T.~Taniguchi, K.~Watanabe,
L.~Jiang and S.~Mizuno for stimulating discussions.  This work was supported
by the WPI Initiative on Materials Nanoarchitectonics, Ministry of Education,
Culture, Sports, Science and Technology of Japan.
\end{acknowledgments}


\begin{thebibliography}{50}
  \bibitem{Wallace1947}
    P. R. Wallace, Phys. Rev. \textbf{71}, 622 (1947).
  \bibitem{Novoselov2004}
    K. S. Novoselov, A. K. Geim, S. V. Morozov, D. Jiang, Y. Zhang, S. V.
    Dubonos, I. V. Grigorieva, and A. A. Firsov, Science \textbf{306}, 666
    (2004).
  \bibitem{GrapheneBook}
    M. I. Katsnelson, \textit{Graphene Carbon in Two dimensions} (Cambridge University
    Press, 2012).
  \bibitem{Geim2009}
     A. K. Geim, Science \textbf{324,} 1530 (2009).
  \bibitem{QHE}
    K. v. Klitzing, G. Dorda, and M. Pepper, Phys. Rev. Lett. \textbf{45}, 494 (1980).
  \bibitem{TKNN}
    D. J. Thouless, M. Kohmoto, M. P. Nightingale, and M. d. Nijs, Phys.
    Rev. Lett. \textbf{49}, 405 (1982).
  \bibitem{Haldane1988}
    F. D. M. Haldane, Phys. Rev. Lett. \textbf{61}, 2015 (1988).
  \bibitem{KaneRMP}
    M. Z. Hasan and C. L. Kane, Rev. Mod. Phys. \textbf{82}, 3045 (2010).
  \bibitem{ZhangRMP}
    X.-L. Qi and S.-C. Zhang, Rev. Mod. Phys. \textbf{83}, 1057 (2011).
  \bibitem{Kane2005a}
    C. L. Kane and E. J. Mele, Phys. Rev. Lett. \textbf{95}, 226801 (2005).
  \bibitem{Kane2005b}
    C. L. Kane and E. J. Mele, Phys. Rev. Lett. \textbf{95}, 146802 (2005).
  \bibitem{GeimNPhys2010}
    F. Guinea, M. I. Katsnelson, and A. K. Geim, Nature Phys. \textbf{6}, 30
    (2010).
  \bibitem{Gomes2012}
    K. K. Gomes, W. Mar, W. Ko, F. Guinea, and H. C. Manoharan, Nature
    \textbf{483}, 306 (2012).
  \bibitem{Hunt2013}
    B. Hunt \textit{et al.}, Science \textbf{340}, 1427 (2013).
  \bibitem{He2013}
    W. Yan, W.-Y. He, Z.-D. Chu, M. Liu, L. Meng, R.-F. Dou, Y. Zhang, Z. Liu,
    J.-C. Nie, and L. He, Nat. Commun. \textbf{4}, 2159 (2013).
  \bibitem{QNiuPRB2010}
    Z. Qiao, S. Yang, W. Feng, W.-K. Tse, J. Ding, Y. Yao, J. Wang, and Q.
    Niu, Phys. Rev. B \textbf{82}, 161414 (2010).
  \bibitem{LiangNJP2013}
    Q.-F. Liang, L.-H. Wu, and X. Hu, New J. Phys. \textbf{15}, 063031 (2013).
  \bibitem{Haldane2008}
    F. D. M. Haldane, and S. Raghu, Phys. Rev. Lett. \textbf{100}, 013904
    (2008).
  \bibitem{Wu2015}
    L.-H. Wu, and X. Hu, Phys. Rev. Lett. \textbf{114}, 223901 (2015).
  \bibitem{Tarruell2012}
    L. Tarruell, D. Greif, T. Uehlinger, G. Jotzu, and T. Esslinger, Nature
    \textbf{483}, 302 (2012).
  \bibitem{FuPRL2011}
    L. Fu, Phys. Rev. Lett. \textbf{106}, 106802 (2011).
  \bibitem{Hsieh2012}
    T. H. Hsieh, H. Lin, J. Liu, W. Duan, A. Bansil, and L. Fu, Nat. Commun.
    \textbf{3}, 982 (2012).
  \bibitem{Dziawa2012}
    P. Dziawa \textit{et al.}, Nature Mater. \textbf{11}, 1023 (2012).
  \bibitem{Xu2012}
    S.-Y. Xu \textit{et al.}, Nat. Commun. \textbf{3}, 1192 (2012).
  \bibitem{Sakurai}
    J. J. Sakurai, \textit{Modern Quantum Mechanics} (Addison Wesley, 1985).
  \bibitem{NetoRMP2009}
    A. H. Castro Neto, F. Guinea, N. M. R. Peres, K. S. Novoselov, and A. K. Geim,
    Rev. Mod. Phys. \textbf{81}, 109 (2009).
  \bibitem{Bernevig2006}
    B. A. Bernevig, T. L. Hughes, and S.-C. Zhang, Science \textbf{314}, 1757 (2006).
  \bibitem{FuPRB2007}
    L. Fu, and C. L. Kane, Phys. Rev. B \textbf{76}, 045302 (2007).
  \bibitem{AndoPRB1991}
    T. Ando, Phys. Rev. B \textbf{44}, 8017 (1991).
  \bibitem{GrothNJP2014}
    C. W. Groth, M. Wimmer, A. R. Akhmerov, and X. Waintal, New J. Phys.
    \textbf{16}, 063065 (2014).
  \bibitem{Landauer1999}
    Y. Imry, and R. Landauer, Rev. Mod. Phys. \textbf{71}, S306 (1999).
  \bibitem{TkachovPRL2010}
    G. Tkachov, and E. M. Hankiewicz, Phys. Rev. Lett. \textbf{104}, 166803
    (2010).
  \bibitem{Konig2007}
    M. K\"{o}nig, S. Wiedmann, C. Br\"{u}ne, A. Roth, H. Buhmann, L. W.
    Molenkamp, X.-L. Qi, and S.-C. Zhang, Science \textbf{318}, 766 (2007).
  \bibitem{Polini2013}
    M. Polini, F. Guinea, M. Lewenstein, H. C. Manoharan, and V. Pellegrini,
    Nature Nanotech. \textbf{8}, 625 (2013).
  \bibitem{Wunsch2008}
    B. Wunsch, F. Guinea, and F. Sols, New J. Phys. \textbf{10}, 103027
    (2008).
  \bibitem{GibertiniPRB2009}
    M. Gibertini, A. Singha, V. Pellegrini, and M. Polini, Phys. Rev. B
    \textbf{79}, 241406 (2009).
  \bibitem{SGLouie2009}
    C.-H. Park, and S. G. Louie, Nano Lett. \textbf{9}, 1793 (2009).
\end{thebibliography}
\end{document}